\documentclass[12pt]{article}
\usepackage{epsfig}
\usepackage{amsmath}

\newcommand{\be}{\begin{equation}}
\newcommand{\ee}{\end{equation}}
\newcommand{\bea}{\begin{eqnarray}}
\newcommand{\eea}{\end{eqnarray}}

\begin{document}

\title{ 
Tetraquark resonances, flip-flop 
\\ and cherry in a broken glass model
}

\author{\\ P. Bicudo, M. Cardoso and N. Cardoso,
\\ CFTP, Dep. F\'{\i}sica, Instituto Superior T\'ecnico,
\\ Av. Rovisco Pais, 1049-001 Lisboa, Portugal}

\maketitle

\begin{center}{\bf Abstract} \end{center}
We develop a formalism to study tetraquarks using the
generalized flip-flop potential, which include the tetraquark
potential component.
Technically this is a difficult problem, needing the solution of the Schr\" odinger
equation in a multidimensional space.
Since the tetraquark may at any time escape to a pair of mesons, here we
study a simplified two-variable toy model and explore the analogy with a cherry
in a glass, but a broken one where the cherry may escape from.
We also compute the decay width in this two-variable picture, solving
the Schr\" odinger equation for the outgoing spherical wave.

\section{Introduction, tetraquarks with flux tubes}

Our main motivation is to contribute to understand whether exotic
hadrons exit or not.
Although there is no QCD theorem ruling out exotics, they are so hard to find,
that many friends even state that either exotics don't exist, or that at least they
should be very broad resonances.
Nevertheless candidates for different continue to exotics exit
\cite{:2009xt}! 
Here we specialize in tetraquarks, the
less difficult multiquarks to compute beyond the baryons and hybrids.
Notice that there are many possible sorts of tetraquarks:
\\ - the borromean 3-hadron molecule
\\ - the Heavy-Heavy-antilight-antilight
\\ - the hybrid-like tetraquark
\\ - the Jaffe-Wilczek diquark-antidiquark with a generalized Fermat string

\vspace{.5cm}
{\bf The borromean 3-hadron molecule}
\\ In an exotic channel, quark exchange leads to repulsion, while quark-antiquark
annihilation is necessary for attraction. A possible way out is adding another meson,
allowing for annihilation, to bind the three body system. This
has already led to the computation of decay widths, which turned out to be wide
\cite{Bicudo:2003rw,LlanesEstrada:2003us}.

\vspace{.5cm}
{\bf The Heavy-Heavy-antilight-antilight}
\\ The heavy quarks are easy to bind since the kinetic energy $p^2 / (2 m)$ is smaller, thus
their Coulomb short distance potential could perhaps provide sufficient binding,
while the light antiquarks would form a cloud around them
\cite{SilvestreBrac:1993ss}.

\vspace{.5cm}
{\bf The hybrid-like tetraquark}
\\ Possibly a quark and antiquark may be in a colour octet, and then the tetraquark
is equivalent to a quark-gluon-antiquark hybrid. Recently we computed in
Lattice QCD the color fields for the static hybrid quark-gluon-antiquark system, 
and studied microscopically the Casimir scaling
\cite{Cardoso:2009kz}.

\begin{figure}[t!]
\begin{center}
\includegraphics[width=.45\linewidth]{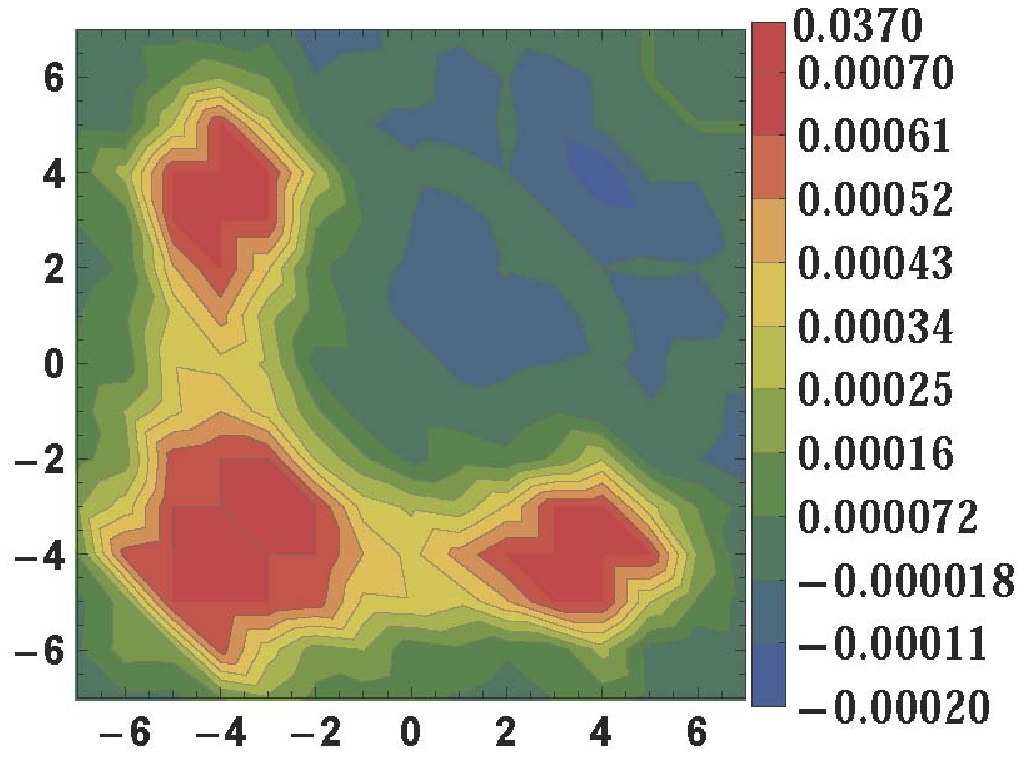}
\includegraphics[width=.5\linewidth]{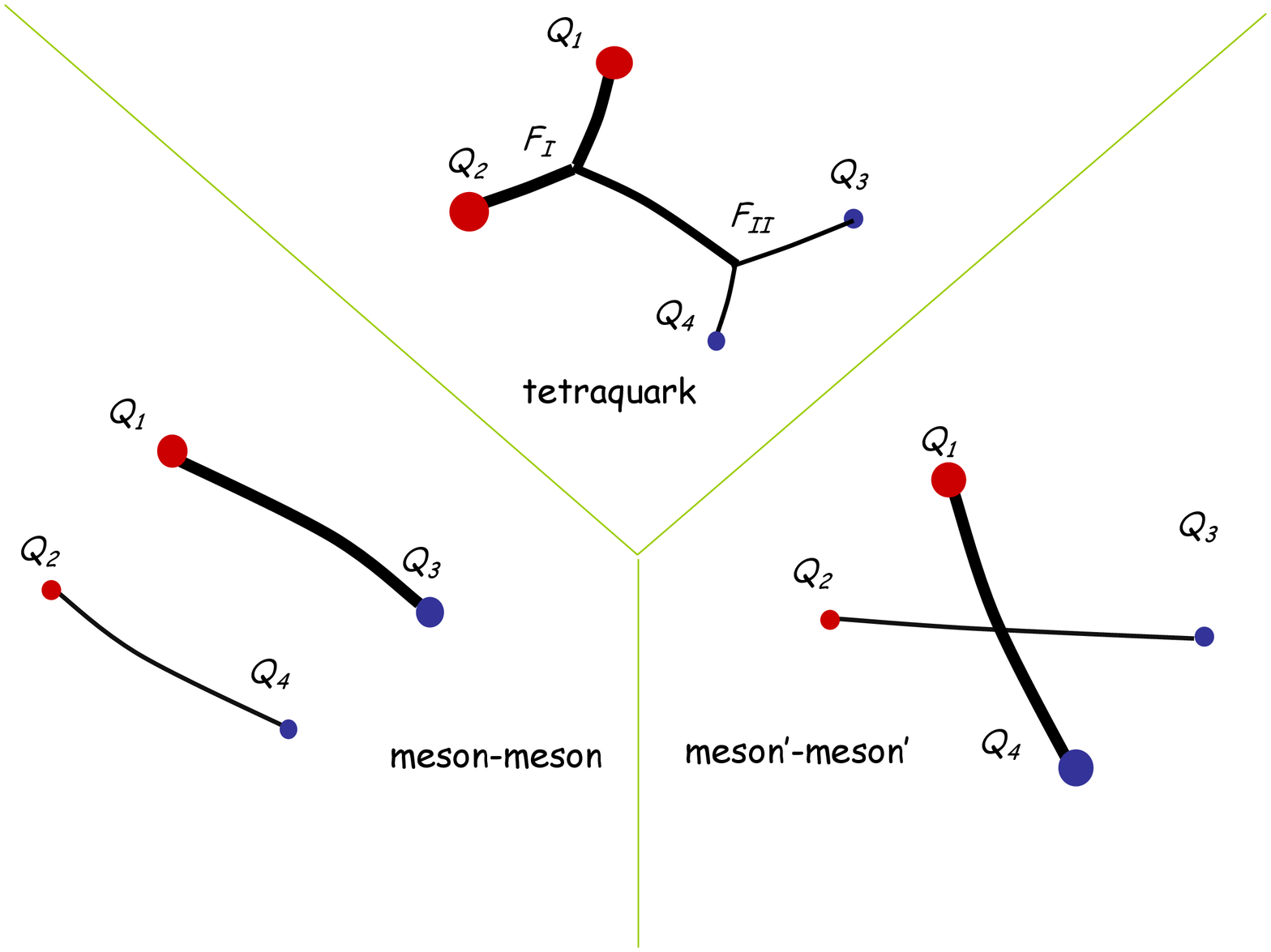}
  \caption{(left) In a hybrid, flux tubes divide into two fundamental flux tunes, one connecting 
the octet with the quark and another connecting the octet to the antiquark. In the
baryon and in the three-gluon glueball, static quenched Lattice QCD simulations also show
confinement via fundamental flux tubes.
(right) Triple flip-flop Potential potential. To the list of
potentials to minimize including usually only two different meson pair potentials, 
we join another potential, the tetraquark potential. 
\label{fluxtubes tripleflipflop}
}
\end{center}
\end{figure}

Notice that our lattice simulation shows that flux tubes prefer to divide
into fundamental flux tubes, or flux tubes carrying a colour triplet flux, 
as we show in Fig. \ref{fluxtubes tripleflipflop}.

\vspace{.5cm}
{\bf The Jaffe-Wilczek diquark-antidiquark with a generalized Fermat string}
\\
Since there is no evidence for long distance polarization forces, or Van der
Waals forces, in hadron-hadron interactions, the two-body confinement
potentials cannot be right for multiquarks
\cite{Gavela:1979zu}!
A solution to this problem consists in considering the flip-flop potential,
where confining flux tubes or strings take the geometry minimizing the
energy of the system. Quark Confinement And Hadronic Interactions
\cite{Lenz:1985jk}.

Again the flux tubes in the tetraquark are expected 
to divide and link into fundamental flux tubes, and a
possible configuration is in a H-like or butterfly-like flux tube. 
This tetraquark can be classified as a Jaffe-Wilczek one
since the quarks are combined in a diquark-like antitriplet
and the antiquarks are combined in a antidiquark-like triplet
\cite{Jaffe}.

The technical difficulty in that framework is to compute the decay widths since
this tetraquark is open for the decay into a pair of mesons. 
Moreover it is expected that the absence of a potential barrier above threshold 
may again produce a very large decay width to any open channel, although
Marek and Lipkin suggested that multiquarks with angular excitations may gain a centrifugal barrier,
leading to narrower decay widths
\cite{Karliner:2003dt}. 

Here we continue a previous work, where
we assumed confined (harmonic oscilator-like) wavefunctions for the confined
objects, one tetraquark and two different pairs of final mesons, and computed
their hamiltonian.
We utilized the Resonating Group Method and were surprised
by finding very small decay widths
\cite{Cardoso:2008dd}.

\vspace{.5cm}
{\bf Our approach to study the tetraquark with a generalized Fermat string}
\\
We thus return to basics and decide to have no overlaps.
We want to solve the Schr\" odinger equation for the four particles, and from the
Schr\" odinger solutions also compute the decay widths. Our starting point is  
the extended triple flip-flop potential
\cite{Vijande}, 
obtained minimizing the three lengths depicted
in Fig.  \ref{fluxtubes tripleflipflop}. 
Recently, we devised a numerical algorithm to compute 
the Fermat points of the tetraquark and the tetraquark potential
\cite{Bicudo:2008yr}.

\begin{figure}[t!]
\begin{center}
  \includegraphics[width=0.55\linewidth]{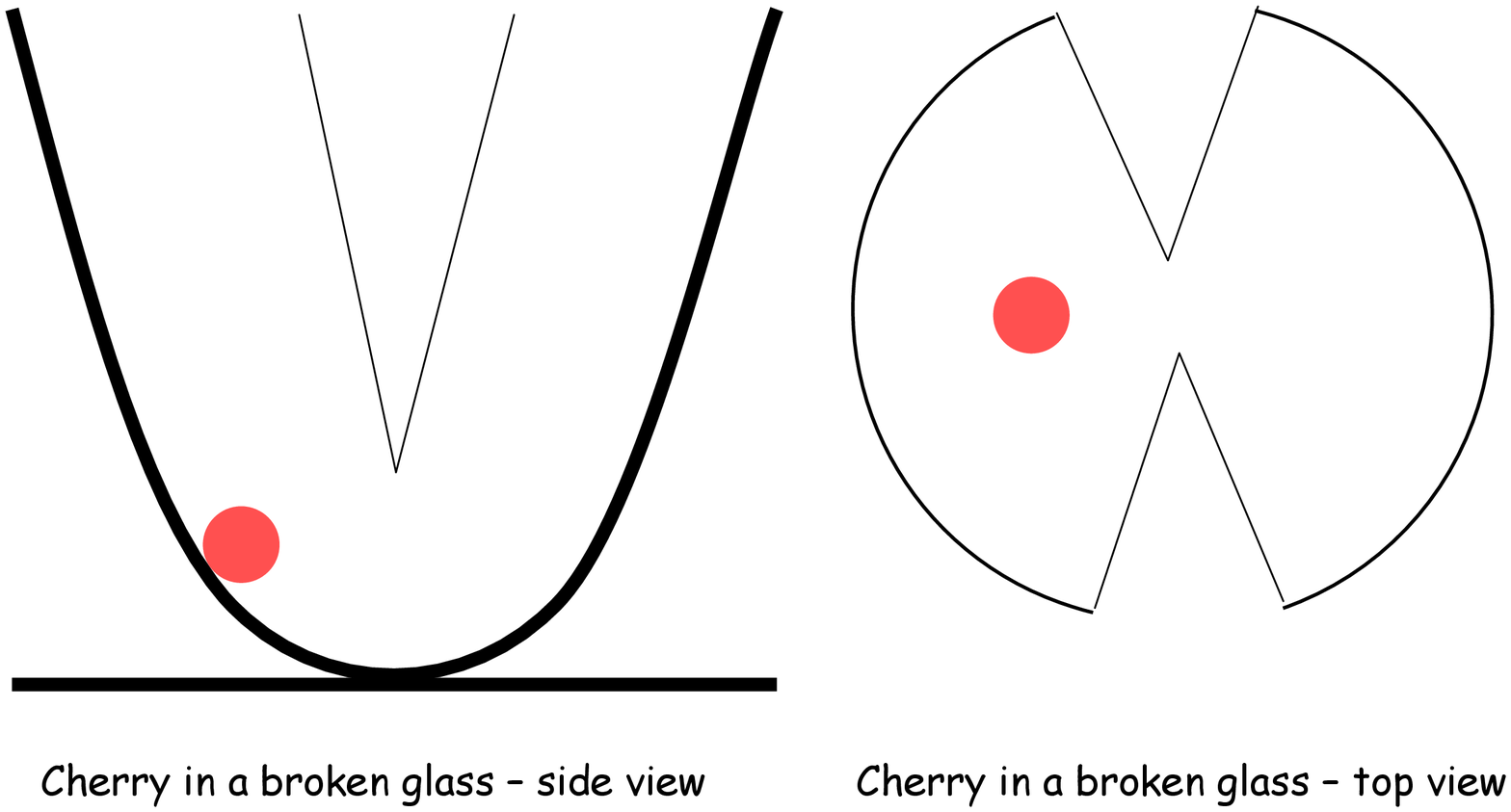}
  \includegraphics[width=0.4\linewidth]{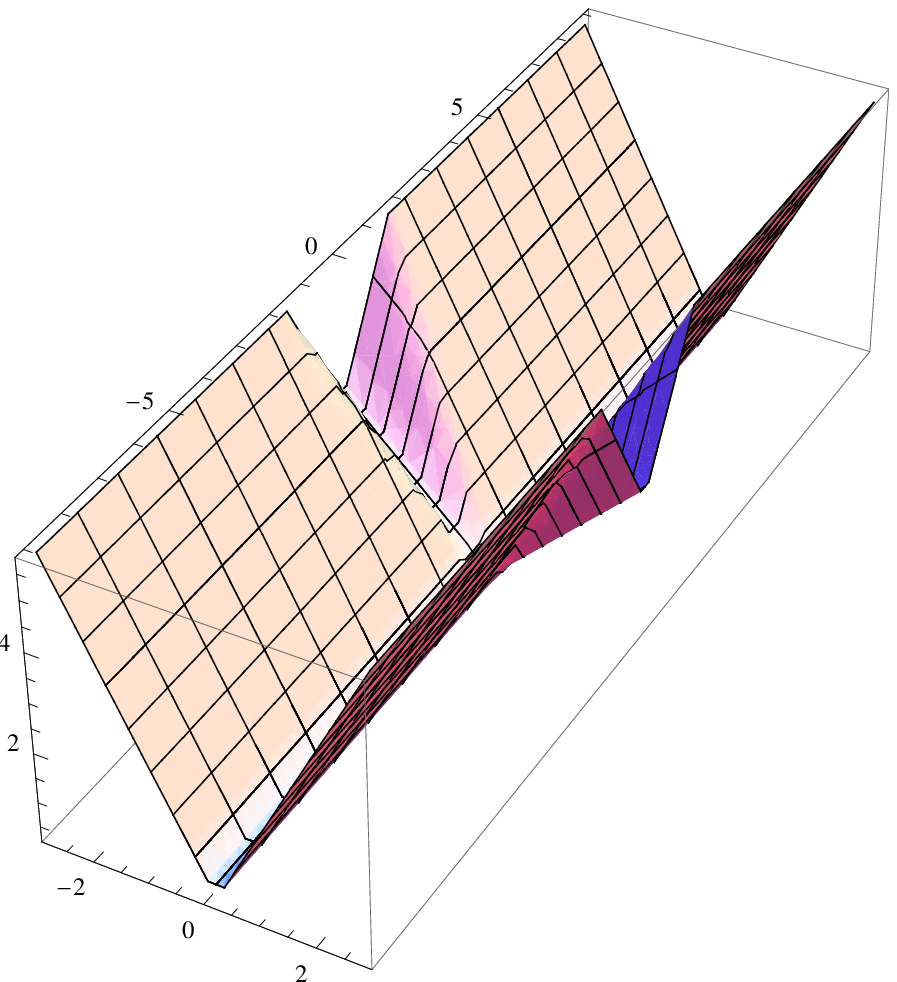}
  \caption{(left) Cherry in a broken glass. Our simplified two-variable toy model is analogous 
to the classical mechanics textbook problem of a cherry in a glass, 
but a broken one where the cherry may escape from. 
Here we solve this model in quantum mechanics, addressing the decay widths of
a system compact in one variable and open in the other. 
(right) Plot of our simplified flip-flop potential, as a function of
the two radial variables $r$ (compact) and $\rho$ (open). 
\label{cherry flip-flop2D}
}
\end{center}
\end{figure}

Solving the Schr\" odinger equation is then a well defined problem which should be solvable, 
placing our system in a large 12 dimensional box.
However this is a very difficult problem. Even assuming s-vaves, we would get 3
variables, some confined and some in the continuum (similar to problems in extra
compactified dimensions or to lattice QCD) so we decide to work in a toy model,
where the number of variables is simplified.
We thus simplify the triple flipflop potential, with a single inter-meson variable,
using the approximation on the diquark and anti-diquark Jacobi coordinates,
\be
\boldsymbol{\rho}_{13}=\boldsymbol{\rho}_{24}
\ee
of having a single internal variable $\boldsymbol{\rho}$ in 
the mesons. We get a flipflop potential where $\rho$ is open to continuum
and $r$ is confined, minimizing only two potentials,
\bea
V_{MM} (r, \rho) & = & \sigma (2 r) \ ,
\\
V_T (r, \rho) &=& \sigma (r + \sqrt 3 \rho) \ .
\eea
Our problem is similar to the classical student's problem of a Cherry in a
glass. 
However this is not a simple student's problem since the glass is broken and the
cherry may escape from the glass! 
The flip-flop and broken glass potentials are depicted in Fig. \ref{cherry flip-flop2D}.
Here we report on our answer
\cite{Bicudo:2010mv}
to the question, 
{\em in the quantum case, are there resonances, and what is their decay width? }

\section{Finite difference method }

\begin{figure}[t!]
\begin{center}
\vspace{1cm}  
  \includegraphics[width=0.45\linewidth]{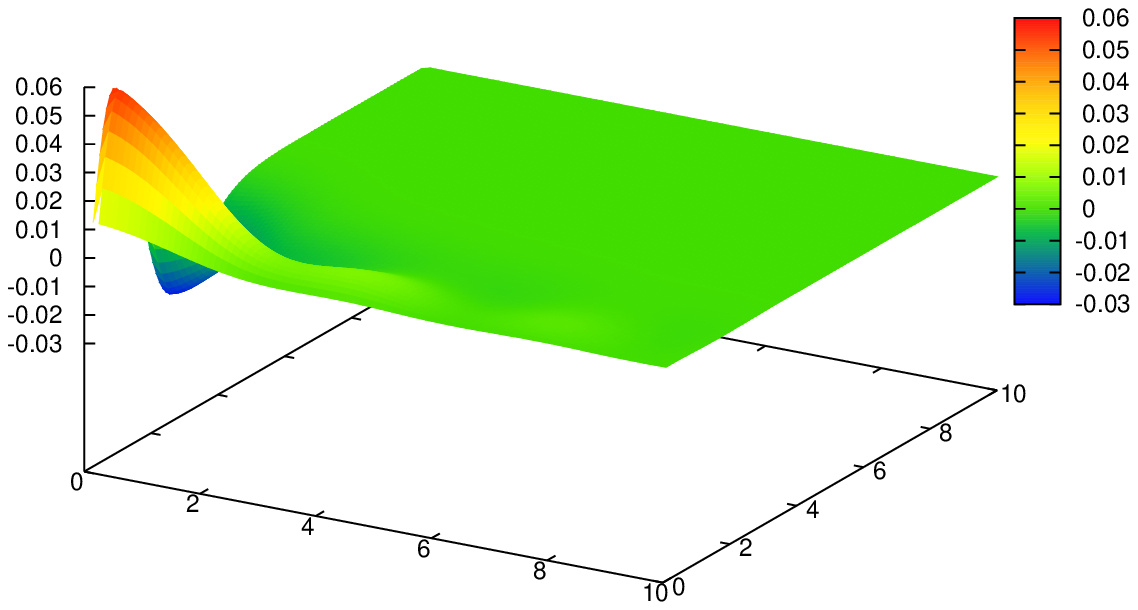}
  \includegraphics[width=0.45\linewidth]{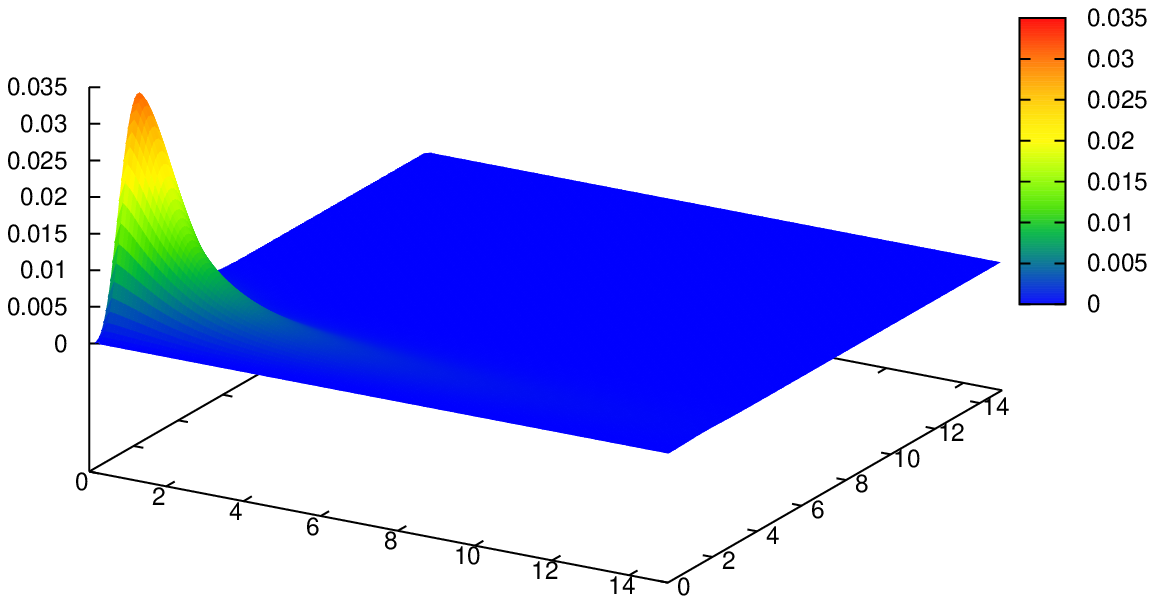}
\caption{(left) Semi-localized state, or resonance for $l_r = 1$.(right) Bound state for $l_r = 3$.
\label{resonance_lr1 boundtstate_lr3}
}
\end{center}
\end{figure}

Since there is a single scale in the potential and a single scale in the kinetic
energy, we can rescale the energy and the coordinates, to get a
dimensionless equation,
\be
H \Phi (r,\rho) = [- \Delta_r /2 - \Delta_\rho /2 + \mbox{min} (r+\sqrt 3 \rho, 2r)] \Phi (r,\rho) = E \Phi (r,\rho) \ ,
\ee
that we first solve with the finite difference method.
Thus, our results and figures are dimensionless.
This case is adequate to study equal mass quarks, where the mesons and the
tetraquark have no constant energy shifts. For instance that would be ok for
the light tetraquark and meson-meson system
\be
u u \bar d \bar d _{(S=2)} \leftrightarrow \rho^+ \ \rho^+ \ ,
\ee
or the heavy quark system
\be
c c \bar c \bar c _{(S=2)} \leftrightarrow J/ \psi \ J/ \psi \, .
\ee

We discretize the space in anisotropic lattices and solve the finite difference
Schr\" odinger equation, in up to 6000 $\times $ 6000 sparse matrices (equivalent to 40
points in the confined direction $\times $ 150 points in the radial continuum direction).
We first look for localized
states, selecting among the
6000 eigenvalues the ones
more concentrated close to
the origin at
$\rho=0$.

\begin{figure}[t!]
\begin{center}
\vspace{0.1cm}  
  \includegraphics[width=0.4\linewidth]{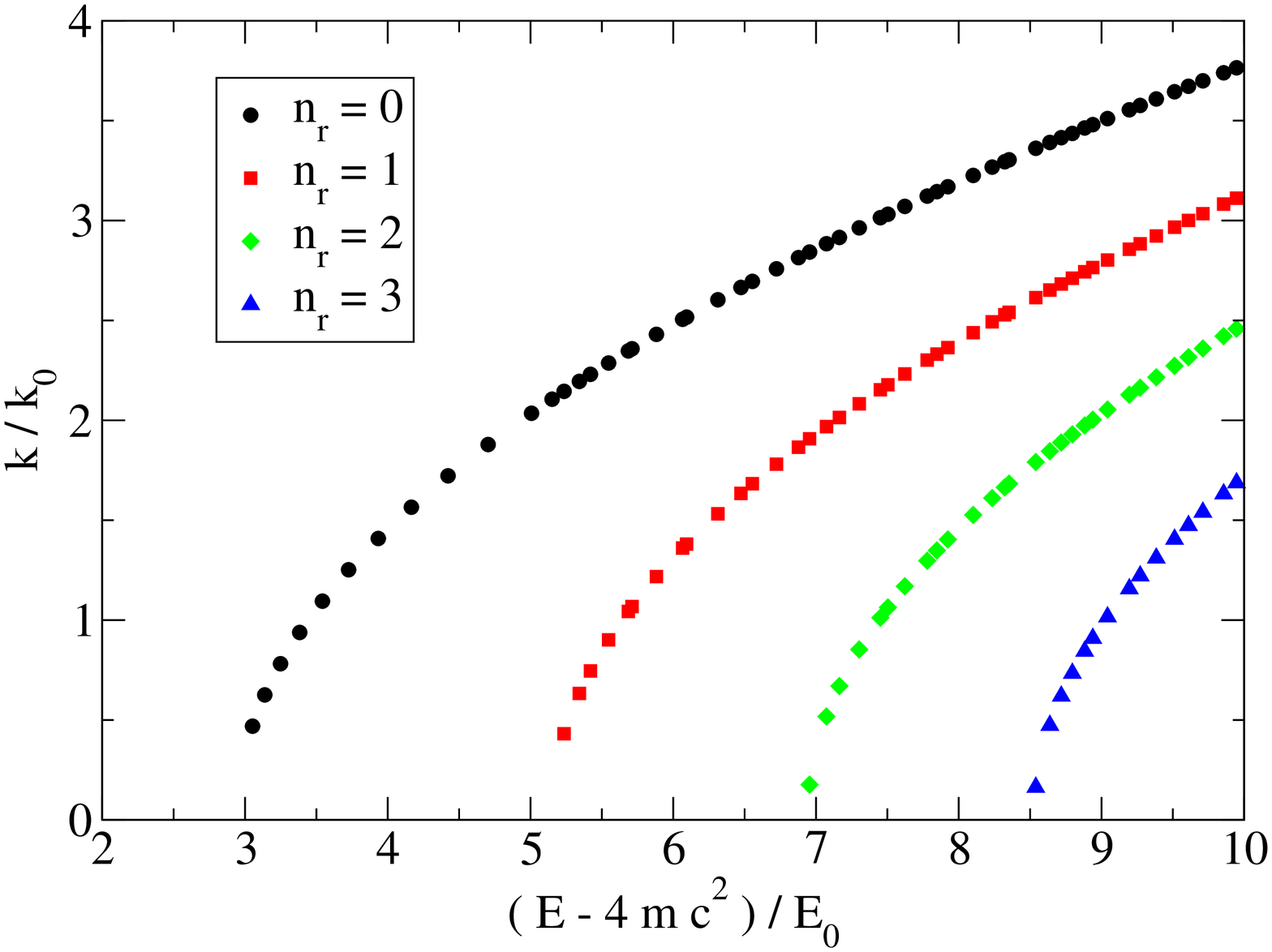}
  \includegraphics[width=0.41\linewidth]{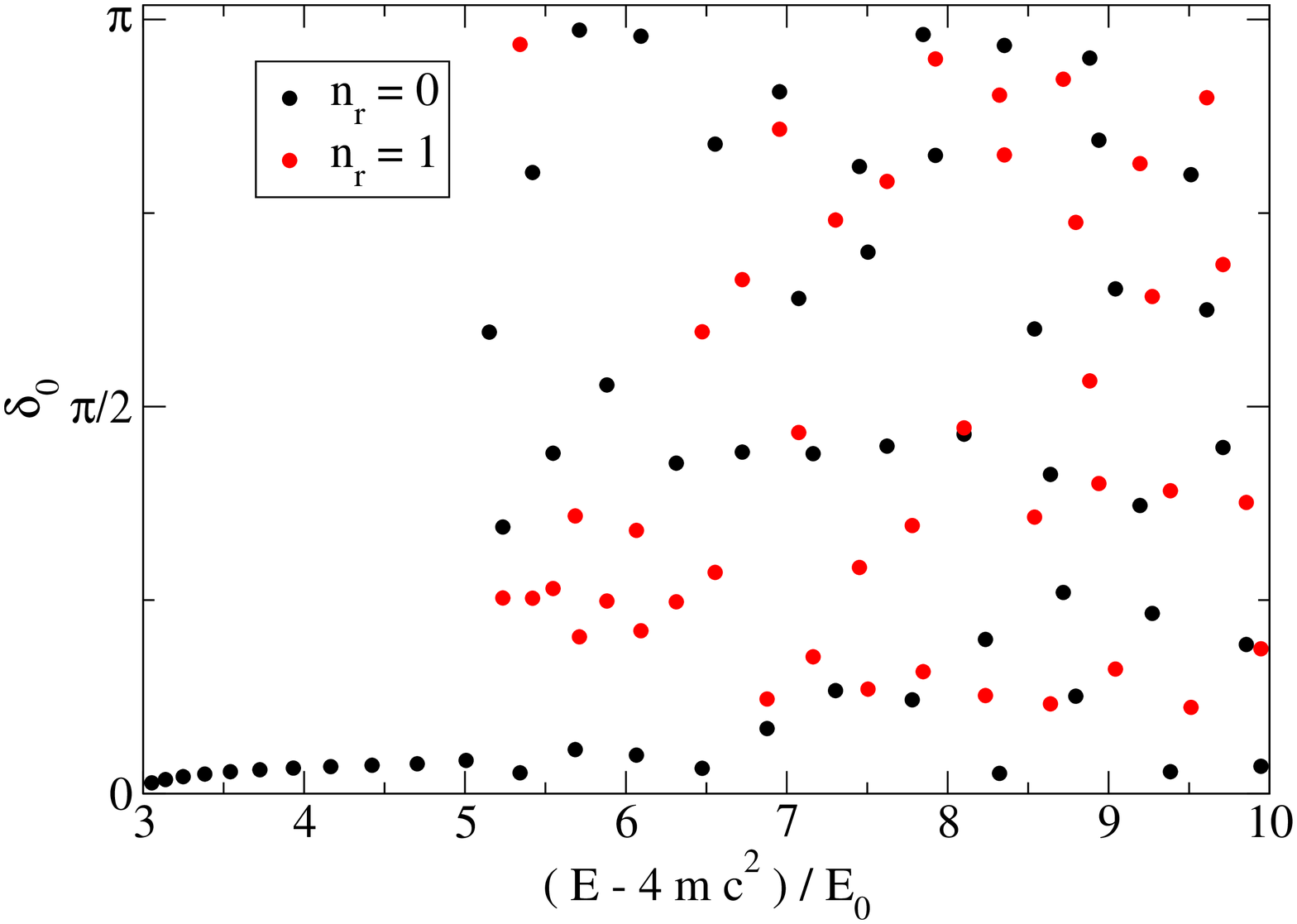}
\caption{  (left) Momenta of the various components as a function of the energy.
(right) "Phase shifts" obtained from the finite differences ( by projecting the eigenstates
in the meson-meson eigenstates ). As can be seen the behaviour is irregular when we have more than one
channel, this is due to the different contributions of multiple channels, for each eigenstate calculated
in the finite difference scheme.
\label{momenta phaseshifts}
}
\end{center}
\end{figure}

To measure the momenta $k_i$ and the phase shifts $\delta_i$, 
we simply fit the large $\rho$ region of the non-vanishing $\psi_i$,
where $i$ indexes the factorized Airy wavefunction in $r$, 
the expression
\be
\psi_i \rightarrow A_i \frac{\sin( k_i \rho + \delta_i )}{ \rho } \ .
\ee
As can be seen in Fig. \ref{momenta phaseshifts}, the momenta $k_i$ obey the relation
\be
k_i(E) = \sqrt{ 2 ( E - \epsilon_i ) } ,
\ee
where $\epsilon_i$ is the threshold energy of the respective channel.

However, the phase shifts we get are not only discrete but 
rather irregular above threshold. In the next Section 3 we compute the
phase shifts with an improved method.

\section{Outgoing spherical wave method}

Because the finite difference method is not entirely satisfactory for the computation
of the phase shifts $\delta$, we move
to another method, consisting in in studying the outgoing spherical waves.
Since the finite difference method shows clearly bands for the different
internal energies of the mesons, we integrate the confined coordinate $r$ with
eigenvalues of the meson equation, i.e. with Airy functions, and thus we are left
with a system of ordinary differential equations in the coordinate $\rho$ .

\vspace{.5cm}
{\bf Projecting onto the $\rho$ coordinate}
\\
We can reduce our problem in the dimensions 
$\rho, \, r$ to a one-dimensional problem in $\rho$ but with of coupled channels. 
We just have to expand the two-dimensional wavefunction as
\be
\Phi( \mathbf{r}, \boldsymbol{\rho} ) = \sum_i \psi_i( \boldsymbol{\rho} ) \phi_i( \mathbf{r} ) \ ,
\label{project}
\ee
where the $\phi_i$ are the eigenfunctions of the $r$ confined hamiltonian.
The one-dimensional potentials $V_{ij}$ are given by
\be
V_{ij}( \rho ) = \int d^3 \mathbf{r} \quad \phi_i^* ( r ) ( V_{FF}( r, \rho ) - V_{MM}(r) )\phi_j ( r )
\ee
where we subtract $V_{MM}$ from the potential, since 
$\hat{H}_{MM}$ is already accounted for in its eigenvalues and eigenfunctions,
used for instance in Eq. (\ref{project}).

\begin{figure}[t!]
\begin{center}
\hspace{-.5cm}
  \includegraphics[width=0.41\linewidth]{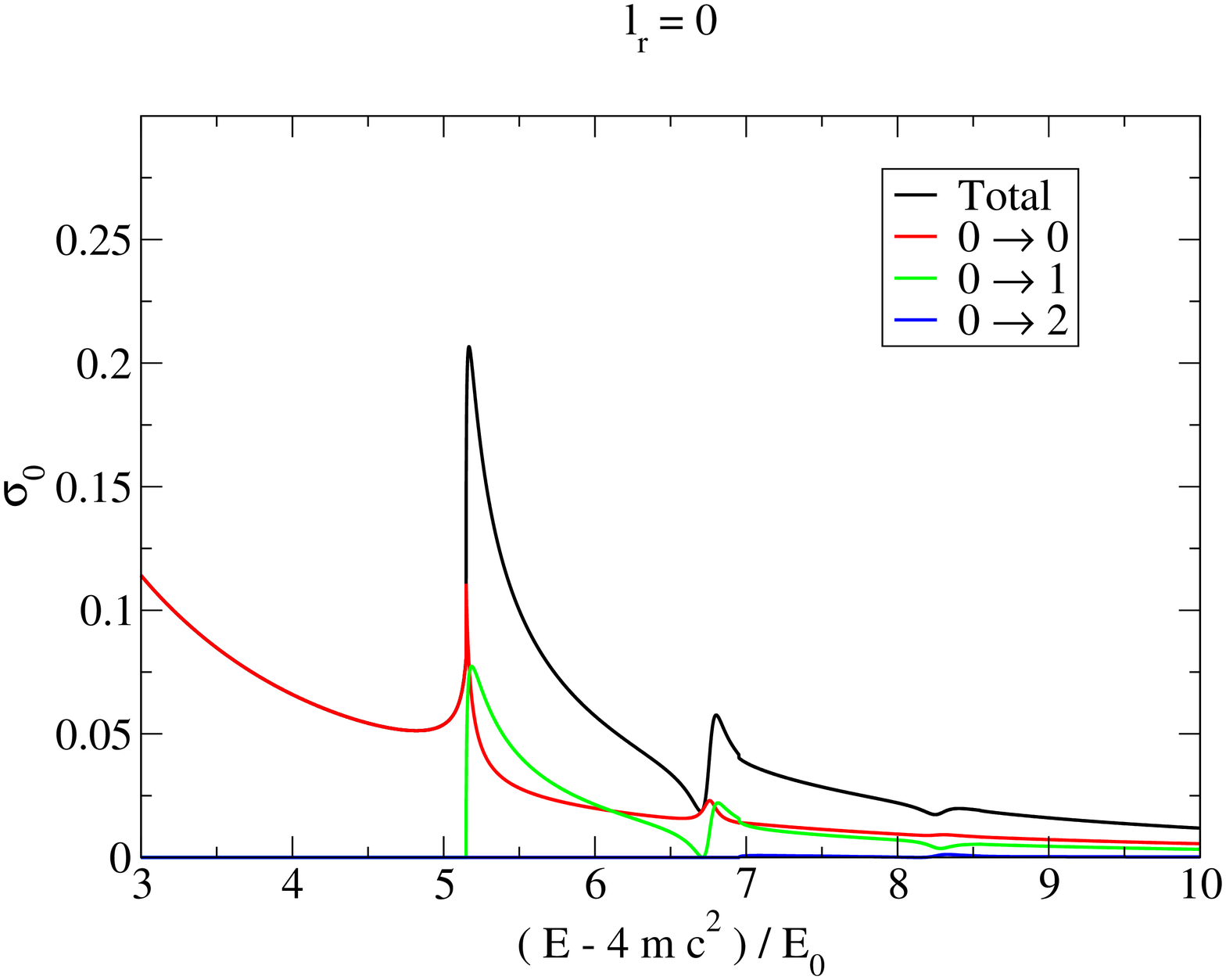}
\hspace{.5cm}
    \includegraphics[width=0.39\linewidth]{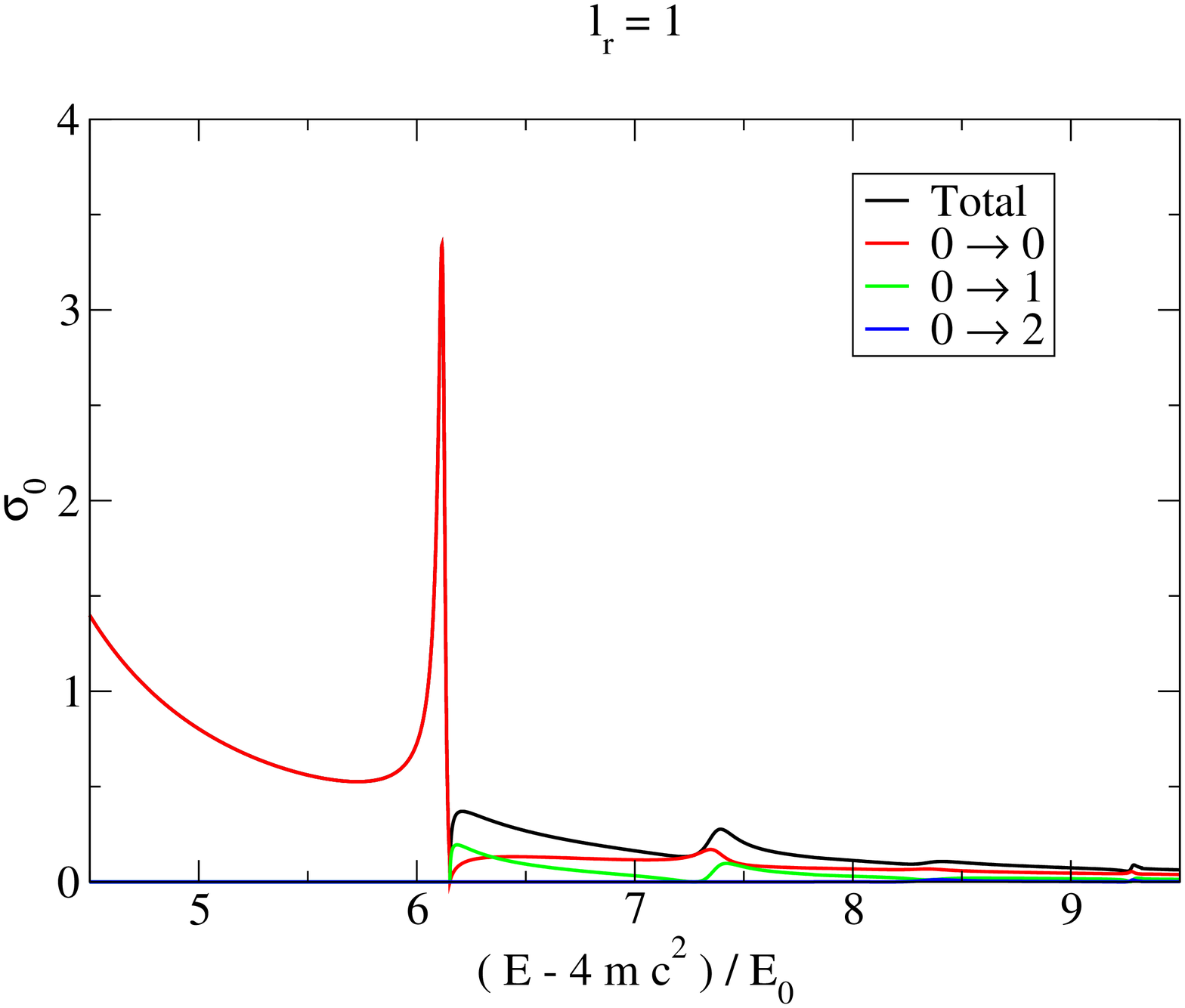}
\caption{ (left) S-wave scattering cross sections from the channel with $l_r = 0$ and $n_r = 0$.
(right)  S-wave scattering cross sections from the channel with $l_r = 1$ and $n_r = 0$.
\label{crosssection_c0_lr0_lrho0 crosssection_c0_lr1_lrho0}
}
\end{center}
\end{figure}

\vspace{.5cm}
{\bf Phase shifts}
\\
We now compute the phase shifts, in order to search for resonances in our simplified flip-flop model.
Solving the outgoing spherical Eq. for this system we can compute the partial cross sections and the total cross section
for the partial wave $l$ --- either directly or by using the optical theorem  ---
and determine the phase shifts as well.

Note that our flip-flop potential has the same scales
of the simple Schr\" odinger equation for a linear potential, which has
a single dimension 
\be
E_0 = \Big( \frac{\hbar^2 \sigma^2}{m} \Big)^{1/3} \,   \ ,
\label{enscale}
\ee
the only energy scale we can construct 
with $\hbar$, $\sigma$ and $m$, the three
relevant constants in the non-relativistic region. 
Thus the number of non-relativistic boundstates or resonances is independent both of the quark mass $m$
and of the string constant $\sigma$.

\vspace{.5cm}
{\bf The centrifugal barrier effect}
\\
Note that we have two distinct angular momenta, which are both conserved, $\mathbf{L}_r = \mathbf{r} \times \mathbf{p}_r$
and $\mathbf{L}_\rho = \boldsymbol{\rho} \times \mathbf{p}_\rho$.
So, each assimptotic state is indexed by its angular momentum $l_r$ and its radial number $n_r$, and the scattering
partial waves are indexed by $l_\rho$.
Thus the system can be diagonalized not only in the scattering angular momenta $\mathbf{L}_\rho$ but also on the confined
angular momenta $\mathbf{L}_r$.
We can describe the scattering process with four quantum numbers: The scattering angular momentum $l_\rho$, the confined
angular momentum $l_r$ and the initial and final states radial number in 
the confined coordinate $r$, $n_i$ and $n_j$.

\begin{figure}
\begin{center}
  \includegraphics[width=0.7\linewidth,height=0.4\linewidth]{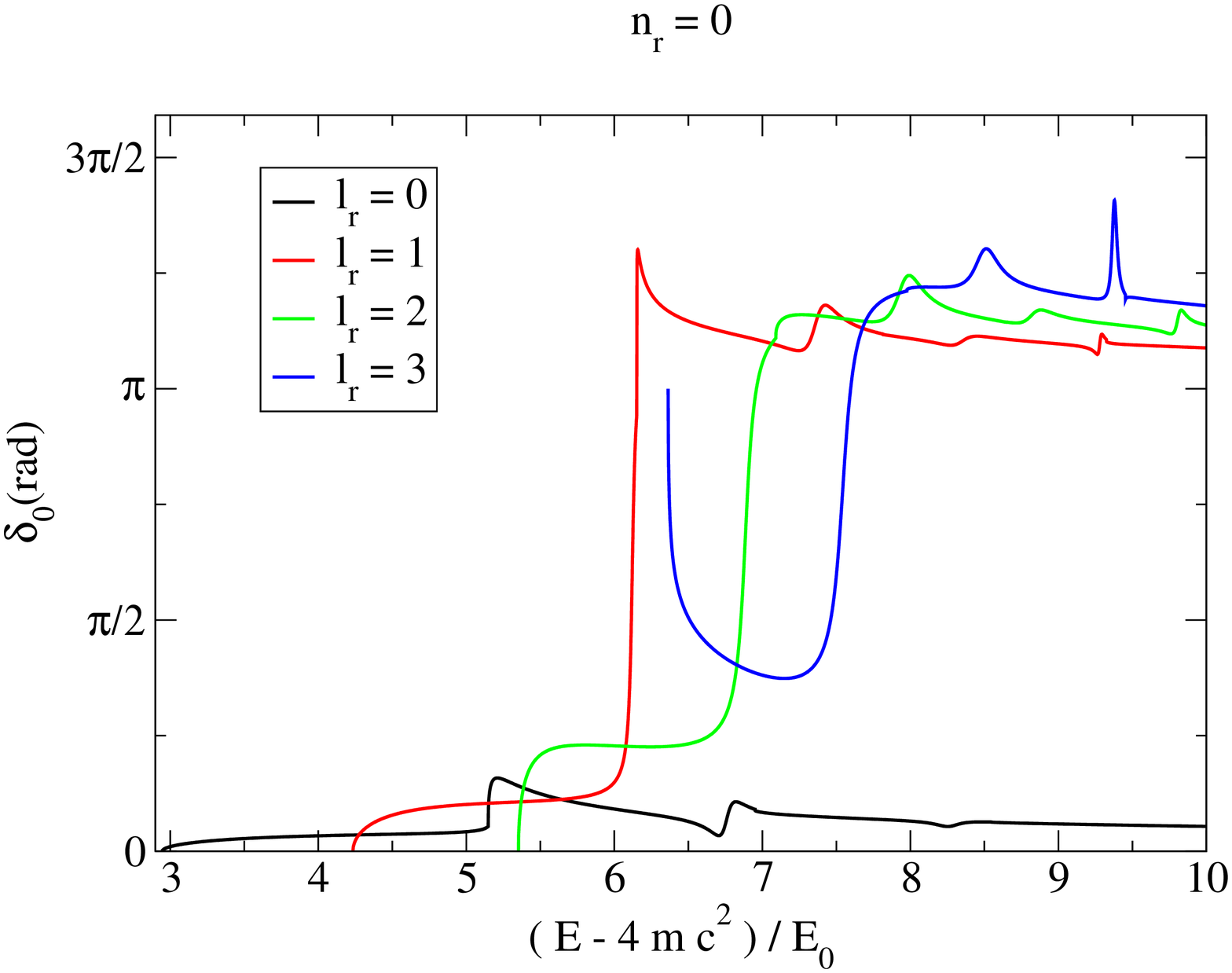}
\caption{ Comparison of the phase shifts for $l_r = 0, 1, 2$ and $3$, with $n_r = 0$.
\label{delta_c0_lrho0}
}
\end{center}
\end{figure}

On Fig. \ref{crosssection_c0_lr0_lrho0 crosssection_c0_lr1_lrho0} we show the $l_\rho = 0$ partial cross sections for the scattering
from the channel with $l_r = 0$ and for $l_r = 1$, with $n_r = 0$. 
Interestingly, the bumps in the cross section seem to occur prior to
the opening of a new channel.

In Fig. \ref{delta_c0_lrho0} we compare the phase shifts for different values of $l_r$,
namely for $l_r = 0, 1, 2$ and $3$.
For $l_r = 0$, we don't observe a resonance, since the phase shift doesn't even cross $\pi/2$.
However, for the $l_r = 1$ and $l_r = 2$ cases, the phase shifts clearly cross the $\pi/2$ line, and a resonance is formed.
This behaviour is somewhat expected, since a centrifugal barrier in $r$ would, in the case of a true tetraquark,
maintain the two diquarks separated, favouring the formation of a bound state.
The tendency of greater stability for greater orbital angular momenta seems to be further confirmed by the
$l_r = 3$, where besides the resonance, a true bound state seems to be formed, as can be seen by the
different qualitative behaviour of the phase shifts for this case.
This bound state formation confirms our observation of a localized states in Section 2, with the finite difference simulation.

Finally we can compute the decay width utilizing the phase shift derivative,
$
\Gamma / 2 = ( d \delta / d E)^{-1}
$
computed when the phase shift $\delta$ crosses $\pi / 2$,
and get the results of Table 	\ref{widthtable1}.
For instance, for light quarks where $m \simeq \sqrt(\sigma) \simeq 400$ MeV
this results in a $l_r=1$ decay width close to 15 MeV.

\section{Conclusion and outlook to tetraquarks}

We study pentaquarks in the Jaffe-Wilczek model,  with a H/butterfly string, but include the open
channels of decays to meson-meson pairs.
We consider an extended flip-flop model, where we add the tetraquark string
to the two-meson strings.
We first apply the RGM method assuming that the mesons have gaussian
wavefunctions, and we obtain very narrow widths.

We then utilize an approximate toy-model, simplifying the number of Jacobi
variables. The model is similar to the model of a Cherry in a Broken Glass.
This allows the solution of the Schr\" odinger equation with finite differences in
a box, where we look for localised states, and try to compute phase shifts.

To compute clearly the phase shifts we then solve the Schr\" odinger equation
for the outgoing spherical waves. We compute de decay widths from the
phase shifts, and we find relatively narrow decay widths. When the
produced mesons are unstable,  the total decay 
width of the tetraquark is then dominated by the final decays of the 
produced mesons.

\begin{table}[t!]
\begin{center}
\begin{tabular}{|c|c|c|}
	\hline
	$l_r$ & $(E - 4 m c^2)/E_0$ & $\Gamma$ / $E_0$ \\
	\hline
	1 & 6.116 & 0.037 \\
	2 & 6.855 & 0.131 \\
	3 & 7.462 & 0.352 \\
	\hline
\end{tabular}
\caption{Decay widths as a function of $l_r$.}
\label{widthtable1}
\end{center}
\end{table}

\vspace{1cm}
\begin{center}{\bf Acknowledgments} \end{center}

We thank George Rupp for useful discussions. We acknowledge the 
financial support of the FCT grants CFTP, CERN/FP/109327/2009 and
CERN/FP/109307/2009.


\end{document}